\documentclass[iop]{emulateapj}

\usepackage{graphicx}  
\usepackage{dcolumn}   
\usepackage{bm}        
\usepackage{amsfonts,amsmath,amssymb,mathrsfs}
\usepackage{color}

\usepackage{hyperref}
\hypersetup{
  colorlinks=true,        
  linkcolor=blue,         
  citecolor=cyan,         
}
\usepackage{natbib,times}
\def\vec#1{\mbox{\boldmath $#1$}}

\newcommand{\cf}{\textit{cf.,}~}
\newcommand{\ie}{\textit{i.e.,}~}
\newcommand{\eg}{\textit{e.g.,}~}

\begin{document}
\shorttitle{Recollimation Shocks in Magnetized Relativistic Jets} 
\shortauthors{Mizuno et al.}
\title{Recollimation Shocks in Magnetized Relativistic Jets}

\author{Yosuke Mizuno\altaffilmark{1, 2}, 
  Jose L. G\'{o}mez\altaffilmark{3}, 
  Ken-Ichi Nishikawa\altaffilmark{4},
  Athina Meli\altaffilmark{5}, 
  Philip E. Hardee\altaffilmark{6}, and
  Luciano Rezzolla\altaffilmark{1, 7} }

\altaffiltext{1}{Institute for Theoretical Physics, Goethe University,
  60438, Frankfurt am Main, Germany; mizuno@th.physik.uni-frankfurt.de}
\altaffiltext{2}{Institute of Astronomy, National Tsing-Hua University,
  Hsinchu 30013, Taiwan, R.O.C.}
\altaffiltext{3}{Instituto de Astrof\'{i}sica de Andaluc\'{i}a (CSIC),
  Apartado 3004, 18080 Granada, Spain}
\altaffiltext{4}{Department of Physics, University of Alabama in
  Huntsville, Huntsville, AL 35899, USA}
\altaffiltext{5}{Department of Physics and Astronomy, University of Gent,
  Proeftuinstraat 86 B-9000, Gent, Belgium}
\altaffiltext{6}{Department of Physics and Astronomy, The University of
  Alabama, Tuscaloosa, AL 35487, USA}
\altaffiltext{7}{Frankfurt Institute for Advanced Studies, Goethe
  University, Ruth-Moufang-Str. 1, 60438 Frankfurt am Main, Germany}

\begin{abstract}
We have performed two-dimensional special-relativistic
magnetohydrodynamic simulations of non-equilibrium over-pressured
relativistic jets in cylindrical geometry. Multiple stationary
recollimation shock and rarefaction structures are produced along the jet
by the nonlinear interaction of shocks and rarefaction waves excited at
the interface between the jet and the surrounding ambient medium.
Although initially the jet is kinematically dominated, we have considered
axial, toroidal and helical magnetic fields to investigate the effects of
different magnetic-field topologies and strengths on the recollimation
structures. We find that an axial field introduces a larger effective
gas-pressure and leads to stronger recollimation shocks and rarefactions,
resulting in larger flow variations. The jet boost grows quadratically with the 
initial magnetic field. On the other hand, a toroidal field
leads to weaker recollimation shocks and rarefactions, modifying
significantly the jet structure after the first recollimation rarefaction
and shock. The jet boost decreases systematically. 
For a helical field, instead, the behaviour depends on the
magnetic pitch, with a phenomenology that ranges between the one seen for
axial and toroidal magnetic fields, respectively. In general, however, a
helical magnetic field yields a more complex shock and rarefaction
substructure close to the inlet that significantly modifies the jet
structure. The differences in shock structure resulting from different
field configurations and strengths may have observable consequences for
disturbances propagating through a stationary recollimation shock.
\end{abstract}

\section{Introduction}

Very Long Baseline Interferometry (VLBI) observations of Active Galactic
Nuclei (AGN) jets often suggests the presence of quasi-stationary
features \citep[see, \eg][]{Jor05, Lis13, Coh14}. These features can be
associated with bends in the jet \citep[see, \eg][]{Alb00} leading to
enhanced emission due to differential Doppler boosting
\citep[see, \eg][]{Gom93}, or to recollimation shocks \citep[see, \eg][]{Dal88,
  Gom95, Gom97, Kom97, Cas13}. Most of the observed quasi-stationary
features appear in the innermost jet regions \citep{Jor05, Jor13, Fro13,
  Coh14} suggesting that they could be associated with recollimation, or
reconfinement shocks produced by a pressure mismatch between the jet and
the external medium.

Multi-wavelength observations of Blazars suggest that high-energy
$\gamma$-ray flares are usually associated with the passing of new
superluminal components through the VLBI core, defined as the bright
compact feature in the upstream end of the jet \citep[see, \eg][]{Mar08,
  Mar10}. In order to produce $\gamma$-ray flares, an increase in
particle and magnetic energy density is required when jet disturbances
cross the radio core. This increase in particle and magnetic density can
naturally be explained by identifying the mm-VLBI radio core with a
recollimation shock \citep{Gom95, Gom97, Mar10, Mar14}. Note however
that at centimeter-wavelengths, the opacity core-shift observed in many
sources \citep[see, \eg][]{Kov08, OSu08, Sok11} suggests that at these
wavelengths the VLBI core corresponds to the transition between the
optically thick-thin regimes.

Recollimation shocks have also been found at hundreds of parsecs from the
central engine in 3C\,120, 3C\,380, and M\,87 \citep{Roc10, Gab14, Asa12,
  Asa14}. The case of M\,87 is particularly interesting since the HST-1
complex, thought associated with a recollimation shock, shows behavior
similar to a VLBI core, with new superluminal components emerging from
its position \citep{Gir12}. Very high-energy emission has also been
observed in connection with variability in the HST-1 region
\citep{Che07} similar to that observed in the VLBI core region of
blazars.

When a jet propagates through an ambient medium, pressure mismatch
between the jet and the ambient medium naturally arises as a result of
ambient pressure decrease. The pressure mismatch drives a radial
oscillating motion of the jet and multiple recollimation regions inside
the jet \citep[see, \eg][]{Gom97, Kom97, Agu01, Alo03, Roc08, Roc09,
  Mim09, Mat12}. The resulting recollimation structure has been
investigated analytically through a self-similar treatment of hydrodynamic
jets \citep{Koh12a} and magnetized jets \citep{Koh12b}, while
\citet{Por14} have studied the causality and stability of magnetized jets
propagating in a decreasing pressure ambient medium. If a significant
rarefaction wave is produced by the recollimation and propagates into the
jet interior, the thermal energy of the plasma can be converted into
kinetic energy, increasing considerably the jet Lorentz factor. This is a
purely relativistic effect, also referred to as the Aloy-Rezzolla (AR)
booster \citep{Alo06}, which takes place in relativistic flows with a
large tangential velocity discontinuity \citep{Rez02}. Under these
conditions, because the quantity $\gamma h$ is conserved across the
rarefaction wave, with $h$ the specific enthalpy and $\gamma$ the Lorentz
factor, large jumps can take place in the latter, leading to a boost
\citep{Rez13}. This boosting mechanism is very basic and has been confirmed
by a number of studies in hydrodynamical and magnetized jets
\citep{Miz08, Kom10, Zen10, Mat12, Sap13}.

In this paper we study in detail how various magnetic field
configurations and strength affect the recollimation-shock structure using
two-dimensional (2D) special relativistic magnetohydrodynamic (RMHD)
simulations. In particular, we focus on the nonlinear rarefactions and
shocks excited at the jet and ambient-medium boundary. Our present 
work provides an
extension of the work presented by \citet{Gom97} to the case of a
dynamically significant magnetic field. 

The paper is organized as follows. In Section 2 we describe the numerical
method and setup used for our simulations. We present our results in
Section 3, illustrating in detail the four different cases of purely
hydrodynamical jets, as well as of jets endowed with axial, toroidal and
helical magnetic fields. Finally we discuss the astrophysical
implications in Section 4, which also contains our conclusions.

\section{Numerical Setup}
\label{sec:ns}
 
We have perform 2D special-relativistic magnetohydrodynamics (SRMHD)
simulations using the three-dimensional general relativistic MHD code
RAISHIN adopting cylindrical coordinates ($R$, $\phi$, $z$) \citep{Miz06,
  Miz11}. In particular, we solve the SRMHD equations in the form 
\begin{widetext}
\begin{eqnarray}
 \partial_t \left(\gamma \rho\right) &+& {1 \over R} \partial_R \left(R
 \gamma \rho v^R\right) + {1 \over R} \partial_\phi \left(\gamma \rho
 v^\phi\right) + \partial_z \left(\gamma \rho v^z\right) = 0\,, \\
 \partial_t S^R &+& {1 \over R} \partial_R \left[R\left(\gamma^2 H
   v^R v^R + p - b^R b^R \right)\right] + {1 \over R} \partial_\phi
 \left(\gamma^2 H v^R v^\phi - b^R b^\phi\right) + \partial_z
 \left(\gamma^2 H v^R v^z - b^R b^z\right) = {\gamma^2 H
   v^\phi v^\phi + p - b^\phi b^\phi \over R}\,, \nonumber \\ \\
  \partial_t S^\phi &+& {1 \over R} \partial_R \left[R \left(\gamma^2
    H v^\phi v^R - b^\phi b^R\right)\right] + {1 \over R}
  \partial_\phi \left(\gamma^2 H v^\phi v^\phi + p - b^\phi
  b^\phi\right) + \partial_z \left(\gamma^2 H v^\phi v^z - b^\phi
  b^z\right) = -{\gamma^2 H v^R v^\phi - b^R b^\phi \over R}\,, \\
 \partial_t S^z &+& {1 \over R} \partial_R \left[R \left(\gamma^2
   H v^z v^R - b^z b^R\right)\right] + {1 \over R} \partial_\phi
 \left(\gamma^2 H v^z v^\phi - b^z b^\phi\right) + \partial_z
 \left(\gamma^2 H v^z v^z + p - b^z b^z\right) = 0\,, \\
\partial_t \tau &+& {1 \over R} \partial_R \left[R \left(\gamma^2
  H v^R - b^R b^t - \gamma \rho v^R\right)\right] + {1 \over R}
\partial_\phi \left(\gamma^2 H v^\phi - b^\phi b^t - \gamma \rho
v^\phi\right) \partial_z \left(\gamma^2 H v^z - b^z b^t - \gamma
\rho v^z\right) = 0\,, \\
 \partial_t B^R &+& {1 \over R} \partial_\phi \left(v^\phi B^R - v^R
 B^\phi\right) + \partial_z \left(v^z B^R - v^R B^z\right) = 0\,,
 \\ \partial_t B^\phi &+& {1 \over R} \partial_r \left[R \left(v^R B^\phi
   - v^\phi B^R\right)\right] + \partial_z \left(v^z B^\phi - v^\phi
 B^z\right) = {v^R B^\phi - v^\phi B^R \over R}\,, \\
\partial_t B^z &+& {1 \over R} \partial_r \left[R \left(v^R B^z - v^z
  B^R\right)\right] + {1 \over R} \partial_\phi \left(v^\phi B^z - v^z
B^\phi\right) = 0\,,
\end{eqnarray}
\end{widetext}
where we have set $c=1$ and used Lorentz-Heaviside units. Here $\rho$ is
the rest-mass density, $\vec{v}$ is the plasma three-velocity, $\vec{S}$
is the three-momentum density, $\tau$ is the conserved energy density, $h
= (e + p_g)/\rho = 1 + \epsilon + p_g/\rho$ is the specific enthalpy,
with $e=\rho (1 + \epsilon)$ the total energy density, $\epsilon$ is the
specific internal energy, and $p_g$ the gas pressure. Furthermore,
$\vec{B}$ is the magnetic field measured in the Eulerian frame, while
$b^\alpha = (b^t, \vec{b})$ is the magnetic field measured in the
comoving frame, so that $b^2 = b^\alpha b_\alpha = B^2/\gamma^2 +
(\vec{v} \cdot \vec{B})^2$, $H \equiv \rho h + b^2$ is the total
enthalpy, $p_m = b^2 / 2$ is the magnetic pressure, and $p = p_g + p_{m}$
the total pressure.  We also adopt as the equation of state that of an ideal
gas with $p_g = \rho \epsilon (\Gamma -1)$ and the adiabatic index
$\Gamma=4/3$ \citep{Rez13}.
 
In the RAISHIN code, a conservative high-resolution shock-capturing
scheme is employed. The numerical fluxes are calculated using the
Harten-Lax-van Leer (HLL) approximate Riemann solver \citep{Har83}, and
the flux-interpolated constrained transport (flux-CT) is used to maintain a
divergence-free magnetic field \citep{Tot00}. We use
the monotonicity-preserving (MP5) spatial interoperation scheme \citep{Sur97}
and the third-order Runge-Kutta time-stepping scheme \citep{Shu88} for all
simulations. The RAISHIN code has the second-order accuracy based on
flux-CT scheme even though we use higher spatial interoperation scheme
\citep{Miz06}. However, the higher-order spatial interoperation scheme
leads to sharper transitions at the discontinuities.

\begin{figure*}
\begin{center}
\includegraphics[height=8.0cm]{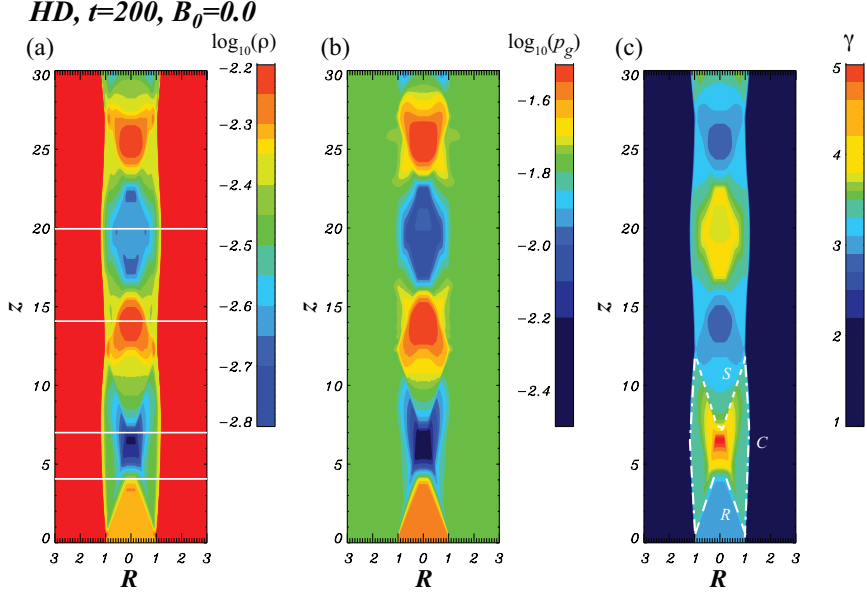}
\end{center}
\caption{Case \texttt{HD}. 2D plots of: ({\it a}) the rest-mass density,
  ({\it b}) the gas pressure, and ({\it c}) the Lorentz factor. The white solid
  lines in panel ({\it a}) indicate the locations along the $z$ axis used
  for the 1D plots in Fig. \ref{1D_hydro}. Shown instead in panel ({\it
    c}) with dashed, dotted and dash-dotted lines are the worldlines of
  the rarefaction waves ($R$), of the the shock waves ($S$), and of the
  contact discontinuities ($C$), respectively.}
\label{2D_hydro}
\end{figure*}

In the simulations presented here, a preexisting cylindrical flow is
established across the simulation domain with jet radius $R_j =1$. This
setup represents a jet far behind the leading-edge Mach disk and bow
shock \citep[see, \eg][]{Miz07, Miz14}. The flow is surrounded by a
higher rest-mass density unmagnetized ambient medium. In all simulations
the rest-mass density ratio is $\eta=\rho_j/\rho_a=5 \times 10^{-3}$,
where the subscripts $j$ and $a$ refer to jet and ambient values,
respectively. The ambient rest-mass density is constant with $\rho_a= 1.0
\,\rho_0$. The jet speed is $v_j=0.9428$ in the $z$ direction and
$\gamma_j = 3$ with local Mach number $M_s=1.69$. We assume that the jet
is initially uniformly over-pressured with $p_{g,j}=1.5\, p_{g,a} =
1.5\, p_{g,0}$ where $p_{g,0}$ is in units of $\rho_0$ \citep[see,
  \eg][]{Gom95, Gom97, Agu01, Mim09}. The gas pressure in the ambient
medium is constant.
   
In order to investigate the effect of the magnetic field on the
recollimation-shock structure, we have considered three different
topologies: axial (or poloidal), toroidal, and helical. More
specifically, for the axial-field case, the initial magnetic field
$\vec{B}$ is uniform and parallel to the jet flow with $B_z=B_0$, while
for the toroidal-field case, we adopt the profile used by \citet{Lin89}
and \citet{Kom99} with
\begin{align}
B_{\phi} = \left\{ 
\begin{array}{llll}
B_0  {R / R_m} &  \qquad \mbox{if\quad }        & R & < R_m\,, \\
B_0 {R_m / R}  & \qquad \mbox{if\quad } R_j \ge & R & \ge R_m\,, \\
0             & \qquad \mbox{if\quad }         & R & > R_j\,,
\end{array}
\right.
\end{align}
where the magnetization radius $R_m=R_j/4$. Finally, for the
helical-field case, we chose a force-free helical magnetic field with
constant magnetic ``pitch'', as the one used in \citet{Miz09, Miz11,
  Miz14}. The poloidal and toroidal magnetic field components are given
by
\begin{equation}
B_{z}= {B_{0} \over 1+ (R/a)^{2}}\,, \qquad
B_{\phi}= {B_{0} (R/a) \over 1+ (R/a)^{2}}\,,
\end{equation}
where $a$ is the characteristic radius of the magnetic field (the
toroidal field component is maximum at radius $a$). The initial magnetic
pitch is defined as
\begin{equation}
\label{eq:mp}
P_0 \equiv \frac{R}{R_j} \left(\frac{B_z}{B_\phi}\right) = \frac{a}{R_j}\,, 
\end{equation}
which is independent of $B_0$ and such that a smaller $P_0$ refers to an
increased magnetic helicity.  In the magnetic helical-field case, we
choose $a = R_j/2$, so that the initial magnetic field has constant helicity and
pitch, with $P_0= 1/2$. The values chosen for the initial $B_0$ and the
corresponding values of the plasma beta parameter, $\beta_p \equiv
p_g/p_m = 2 p_g/b^2$ (averaged and local minimum), as well as the
magnetization parameter $\sigma \equiv b^2/\rho h$ (averaged and local
maximum) are listed in Table 1 for each model. Note that in all cases the
relativistic jet is kinematically dominated for all values of $B_0$.

\begin{figure*}
\begin{center}
\includegraphics[width=0.75\textwidth]{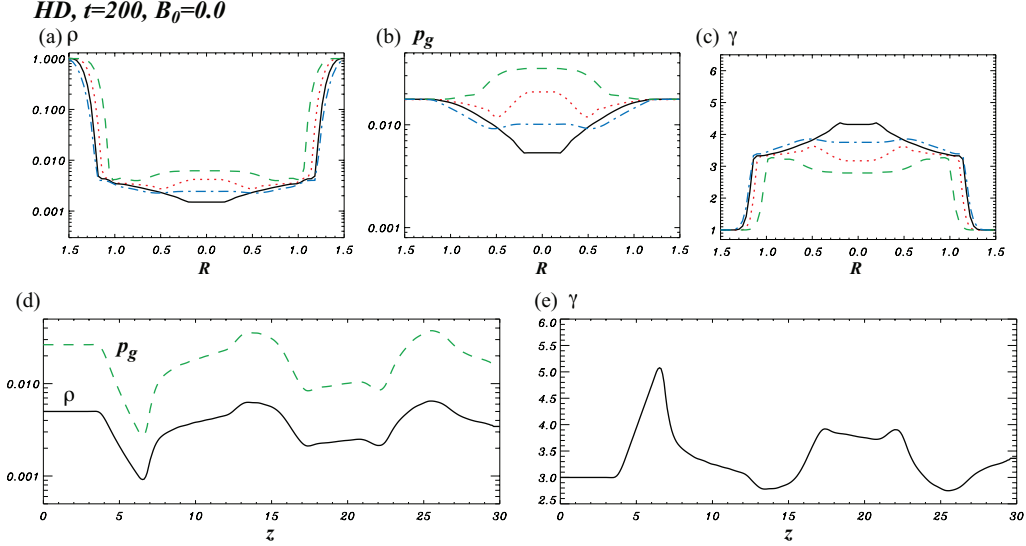}
\end{center}
\caption{{\it Upper row:} 1D profiles perpendicular to jet axis of:
  ({\it a}) the rest-mass density, ({\it b}) the gas pressure, and ({\it c})
  the Lorentz factor as measured at $z=4$ (red dotted), $7$ (black solid),
  $14$ (green dashed), and $20$ (blue dash-dotted), respectively. {\it
    Lower row:} 1D profiles along the jet axis ($R=0$) of: ({\it d})
  the rest-mass density (solid) and the gas pressure (green dashed), and ({\it
    e}) the Lorentz factor. All panels refer to $t_s =
  200$. \label{1D_hydro}}
\end{figure*}

\begin{table}
\begin{tabular}{lcccccc}
\hline
\hline
{Case} & {$B$-field} & {$B_0$} & {$\langle \beta_{p} \rangle$} & {$\beta_{p,{\rm min}}$} & {$\langle \sigma \rangle$} & {$\sigma_{\rm max}$}\\
\hline
\texttt{HD}   & --       & $0.0$ & --    & --    & --                  & --\\
\texttt{\texttt{MHD-a}} & axial    & $0.1$ & $6.2$ & $5.3$ & $8.4 \times 10^{-2}$ & $9.0\times 10^{-2}$ \\  
\texttt{\texttt{MHD-b}} & toroidal & $0.2$ & $140.5$ & $13.3$ & $1.1 \times 10^{-2}$ & $3.6 \times 10^{-2}$ \\
\texttt{\texttt{MHD-c}} & helical  & $0.2$ & $10.8$ & $1.3$ & $1.4 \times 10^{-1}$ & $3.6 \times 10^{-1}$ \\
\hline
\end{tabular}
\caption{Basic properties of the various cases simulated. Listed are the
  magnetic-field topology, the initial magnetic-field strength $B_0$, the
  average and minimum plasma beta parameters $\langle \beta_{p} \rangle$
  and $\beta_{p, {\rm min}}$, as well as the average and maximum
  magnetization parameters $\langle \sigma \rangle$ and $\sigma_{\rm
    max}$.}
\label{table1}
\end{table}

The computational domain is $5\,R_j \times 30\,R_j$ with a uniform grid
of $(R, z)=(128, 300)$ computational zones. We impose outflow boundary
conditions on the surfaces at $R=R_{\rm max}$ and $z=z_{\rm max}$. At $z=0$ we
use fixed boundary conditions and continuously inject the over-pressured
jet into the computational domain. The axisymmetry implies reflecting
boundary conditions at $R=0$.
 
\section{Results}

In what follows we present the results of the simulations going through
the four different magnetic-field configurations considered, \ie cases
\texttt{HD}, \texttt{MHD-a}, \texttt{MHD-b}, and \texttt{MHD-c}.

\subsection{Purely hydrodynamic jet}
\label{sec:hd}
 
We start our discussion of the results by illustrating what can be
considered our reference configuration, that is, a purely hydrodynamical
jet (case \texttt{HD}). Figure \ref{2D_hydro} shows 2D plots of the
rest-mass density, gas pressure and Lorentz factor for the hydrodynamic
case at $t_s=200$, where $t_s$ is in units of $R_j$. Multiple
recollimation shocks and rarefactions are evident along the jet
propagation direction. Downstream of the inlet the over-pressured jet
produces initially a weak conical shock that propagates into the ambient
medium and a quasi-stationary conical rarefaction wave that propagates
into the jet (dashed lines in panel (\textit{c}) of
Fig. \ref{2D_hydro}). The initial shock is rather weak because of the
small initial pressure discontinuity and leads to relatively smooth
transition between jet and ambient medium. The over-pressured jet expands
(following contact discontinuities, dash-dotted lines in
Fig. \ref{2D_hydro}) and the rarefaction wave leads to the conversion of
thermal energy into kinetic energy of the jet \citep[see, \eg][]{Alo06,
  Mat12}. The resulting acceleration is however rather modest given the
choice made here for the initial conditions.

One-dimensional (1D) profiles of hydrodynamic quantities perpendicular to
the jet at axial positions $z=4$, $7$, $14$, and $20$ are shown in the
upper row of Fig. \ref{1D_hydro} and help interpret the dynamics of the
recollimation shocks. Note that at $z \simeq 4$ the conical rarefaction
wave converges to the jet axis and at this point the rest-mass density
and pressure drop significantly and the flow is considerably accelerated,
going from $\gamma \sim 3$ to $\sim 4.6$ (see the solid lines in
Fig. \ref{1D_hydro}). At the same time, the convergence of the rarefaction
wave towards the jet axis leads to strong gas-pressure gradients, which
slow jet expansion. The radial expansion of the jet ceases at $z \simeq
7$ and at $z \gtrsim 7$ the jet starts to contract. Beyond $z \simeq 8$,
a conical shock wave moves outward from the jet axis (see dotted lines in
the panel (\textit{c}) of Fig. \ref{2D_hydro}) and reaches the jet edge
at $z \simeq 13$. At the conical shock, between $z \simeq 8 - 13$, the
rest-mass density and gas pressure increase and the jet Lorentz factor
decreases. Furthermore, when the shock encounters the contact
discontinuity at $z \simeq 13$, the jet structure has mostly returned to
the initial structure, albeit at slightly higher pressure and lower
velocity in the jet. The innermost recollimation structures become
stationary within several light crossing times.

The panels in the bottom row of Fig. \ref{1D_hydro} offer a complementary
view and, in particular, they show 1D profiles of hydrodynamical
quantities along the jet axis, \ie at $R=0$. Note that, as one would
expect, the pressure profile follows exactly that of the rest-mass
density as a result of the ideal-gas equation of state employed. The
Lorentz factor, on the other hand, is inversely proportional to the
rest-mass density, so that the energy conversion from thermal to kinetic
operated by the rarefaction wave leads to a maximum in the Lorentz factor
there where the rest-mass density has a minimum. Additional accelerations
(\ie peaks in the Lorentz factor) can be seen at $z \simeq 17-22$, but
these are less pronounced as the second set of rarefaction waves are not
as efficient in reducing the rest-mass density locally. Note also that
after the first cycle of recollimation shocks and rarefaction waves, the
jet does not return to the initial conditions and so the subsequent
recollimation shocks and rarefactions are weaker overall.

\begin{figure*}
\begin{center}
\includegraphics[height=7.0cm]{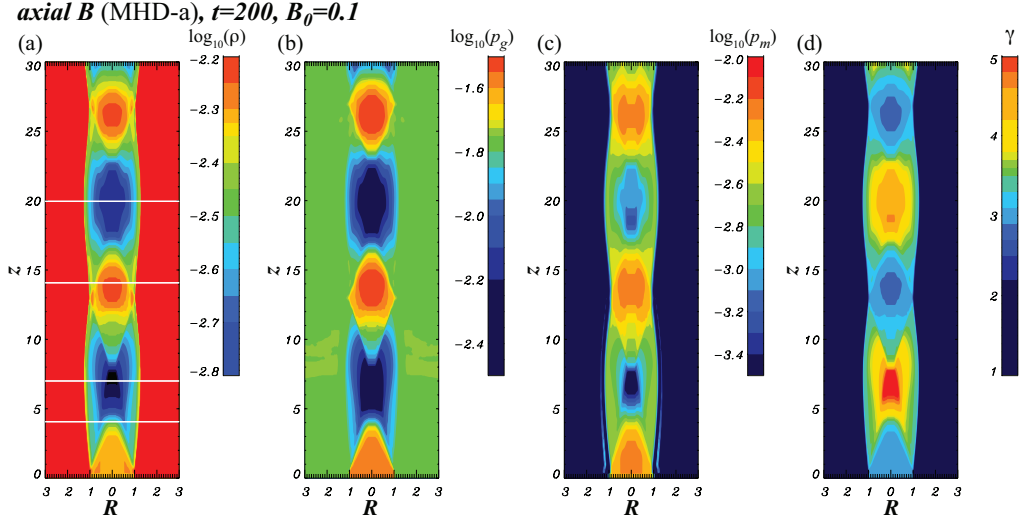}
\end{center}
\caption{Case \texttt{MHD-a}. 2D plots of: ({\it a}) the rest-mass
  density, ({\it b}) the gas pressure, ({\it c}) the magnetic pressure, and ({\it
    d}) the Lorentz factor for the axial magnetic-field case with $B_0
  =0.1$. The white solid lines in panel ({\it a}) indicate the locations
  along the $z$ axis used for the 1D plots in Fig. \ref{1D_axial}.  All
  panels refer to $t_s = 200$. 
  The white solid lines in panel ({\it a}) indicate the 1D plot locations 
  in Fig. \ref{1D_axial}.
  \label{2D_axial}}
\end{figure*}

\subsection{Axial magnetic field}
\label{sec:mhd-a}

Next, we consider the dynamics of the recollimation shocks when an axial
magnetic field is present initially (case \texttt{MHD-a}). Figure
\ref{2D_axial} shows 2D plots of the rest-mass density, gas pressure,
Lorentz factor and magnetic pressure for the axial field case with
$B_0=0.1$ (\texttt{MHD-a}) at $t_s=200$. 

\begin{figure*}
\begin{center}
\includegraphics[width=0.75\textwidth]{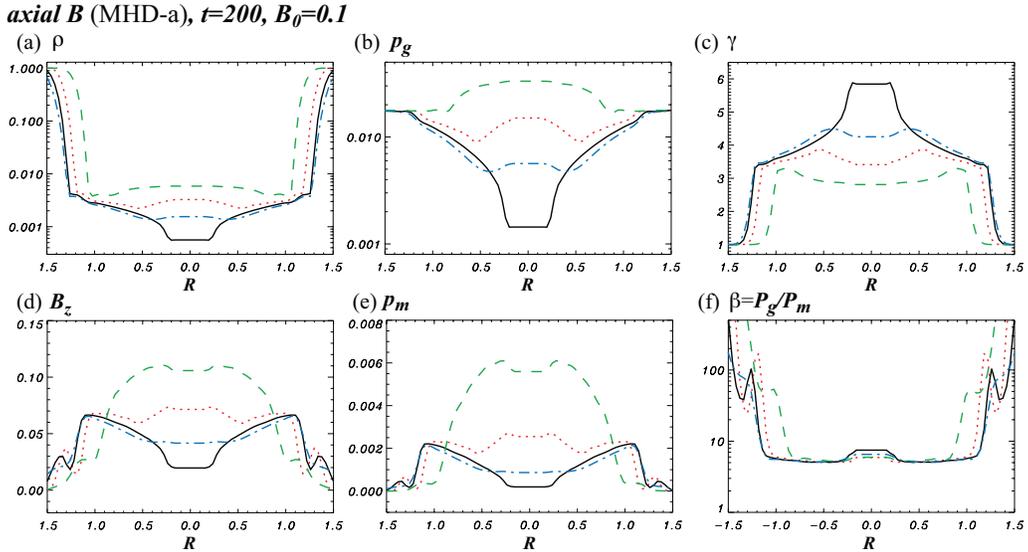}
\end{center}
\caption{1D profiles perpendicular to jet axis of: ({\it a})
  the rest-mass density, ({\it b}) the gas pressure, ({\it c}) the Lorentz factor,
  ({\it d}) the axial magnetic field, ({\it e}) the magnetic pressure, and ({\it
    f}) the plasma beta. Different lines refer to $z=4$ (black solid), $7$
  (red dotted), $14$ (green dashed), and $20$ (blue dash-dotted). All
  quantities refer for the axial magnetic-field case (\texttt{MHD-a})
  with $B_0=0.1$ and $t_s=200$. \label{1D_axial}}
\end{figure*}

As in the hydrodynamic case discussed in the previous Section,
immediately downstream of the inlet, the over-pressured axially
magnetized jet produces a weak conical shock that propagates into the
ambient medium and a quasi-stationary conical rarefaction wave that
propagates into the jet. Also in this case, the flow is boosted as a
result of the exchange from thermal energy to kinetic energy. More
precisely, at $z \simeq 4.5$ the conical rarefaction wave converges to
the jet axis and the rest-mass density, the gas- and the magnetic
pressures drop significantly, leading to an acceleration of the flow from
$\gamma \sim 3$ to $\sim 6.0$ (see the solid lines in
Fig. \ref{1D_axial}). 

Note that the acceleration is \textit{larger} than in the hydrodynamic
case because the rarefaction wave is stronger and the latter is stronger
because the axial magnetic field pressure acts in concert with the gas
pressure but with different dependence on the jet radius. Similar results
were obtained in 2D planar simulations of a flow bounded by a lower
pressured ambient medium \citep{Miz08,Zen10}. Furthermore, downstream of
the rarefaction the gas pressure decreases more than the magnetic
pressure and the plasma beta decreases.

\begin{figure}
\begin{center}
\includegraphics[height=10.0cm]{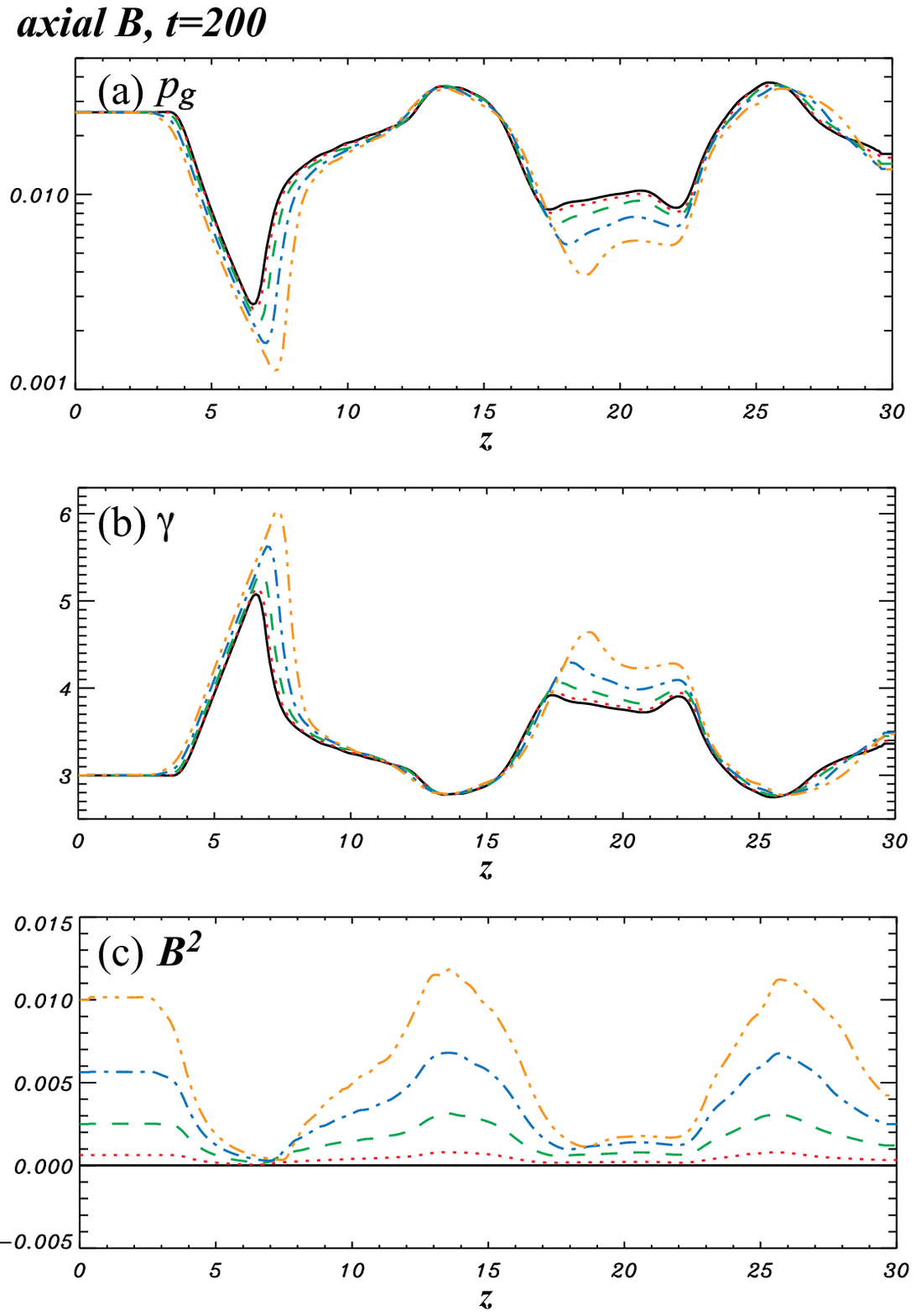}
\end{center}
\caption{1D profiles along the jet axis ($R=0$) of: ({\it a}) the gas
  pressure, ({\it b}) the Lorentz factor, and ({\it c}) the magnetic pressure for
  the axial magnetic-field case. Different lines refer to different
  values of the initial magnetic-field strength, \ie $B_0=0$ (black
  solid), $0.025$ (red dotted), $0.05$ (green dashed), $0.075$ (blue
  dash-dotted), and $0.1$, respectively (orange dash-double-dotted). All
  panels refer to $t_s = 200$. \label{1D_onz_axial_comp}}
\end{figure}

Figure \ref{1D_axial} reports 1D profiles of various physical quantities
perpendicular to the jet axis at axial positions $z=4$, $7$,
$14$, and $20$ [see white solid lines in panel (\textit{a}) of
  Fig. \ref{2D_axial}].  In this way it is possible to appreciate that
the convergence of the rarefaction wave towards the jet axis leads to
strong pressure gradients (both gas and magnetic), which slow jet
expansion. Indeed, the radial expansion ceases at $z \simeq 7$ and at $z
\gtrsim 7$ the jet starts to contract. Beyond $z \simeq 8$, a conical
shock wave moves outward from the jet axis and reaches the jet edge at $z
\simeq 13$. When the shock encounters the contact discontinuity at $z
\simeq 13$, the jet has mostly returned to the initial structure, albeit
at a slightly lower velocity.

In summary, while maintaining many similarities with the purely
hydrodynamical evolution, the presence of an axial magnetic field leads
to stronger recollimation shock and rarefaction waves. In turn, because
the AR booster is sensitive only to jumps in the specific enthalpy and
not on whether a magnetic field is present, the sharper discontinuities
in the flow lead to stronger boosts (see also the discussion in
Sect. \ref{se:ja}).

Finally, to assess the role played by the initial strength of the
magnetic field $B_0$ on the subsequent dynamics, we show in
Fig. \ref{1D_onz_axial_comp} the 1D profiles of the gas pressure, of the
Lorentz factor and of the magnetic-field strength along the jet axis, \ie
at $R = 0$. The different lines refer respectively to $B_0=0.0$, \ie the
purely hydrodynamical evolution, as well as $B_0=0.025$, $0.05$, $0.075$,
and $0.1$ respectively. The main feature to note in this case is that an
increasingly stronger axial magnetic field does not introduce
qualitative changes in the flow dynamics. Indeed, the changes are only
quantitative, with smaller values for the rest-mass density and pressure
and consequently larger values of the Lorentz factor. It follows that the
observational knowledge of the plasma velocity at different positions in
the jet could be used to deduce the strength of the magnetic field in the
jet.

\begin{figure*}
\begin{center}
\includegraphics[height=7.0cm]{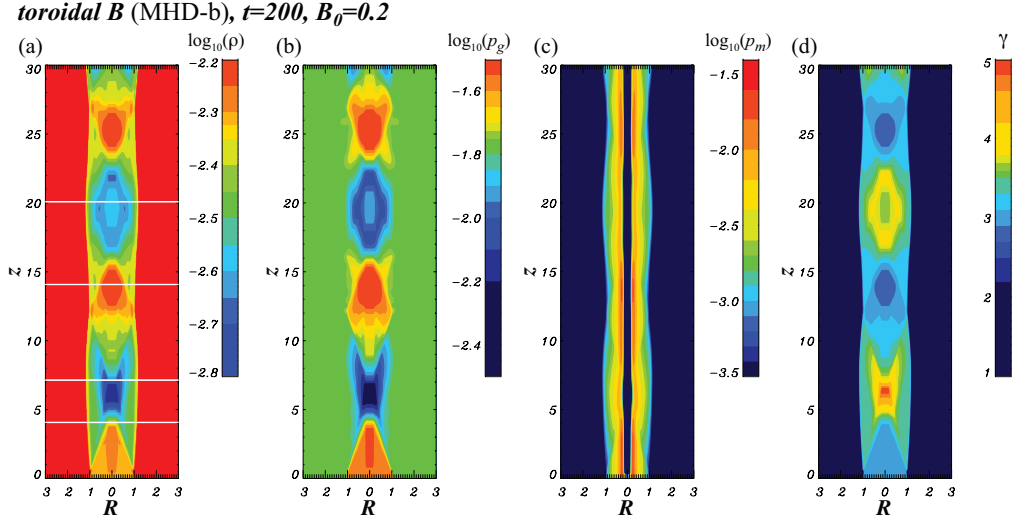}
\end{center}
\caption{Case \texttt{MHD-b}. The same as in Fig. \ref{2D_axial} but for
  the toroidal magnetic-field case with $B_0 =0.2$. \label{2D_toroidal}}
\end{figure*}

\subsection{Toroidal magnetic field}
\label{sec:mhd-b}

We continue our investigation by considering the dynamics of the
recollimation shocks when a toroidal magnetic field is initially present
(case \texttt{MHD-b}). In particular, Fig. \ref{2D_toroidal}
shows 2D plots of the rest-mass density, gas pressure, magnetic pressure
and Lorentz factor for the toroidal field case with $B_0=0.2$ at
$t_s=200$.

\begin{figure*}
\begin{center}
\includegraphics[width=0.75\textwidth]{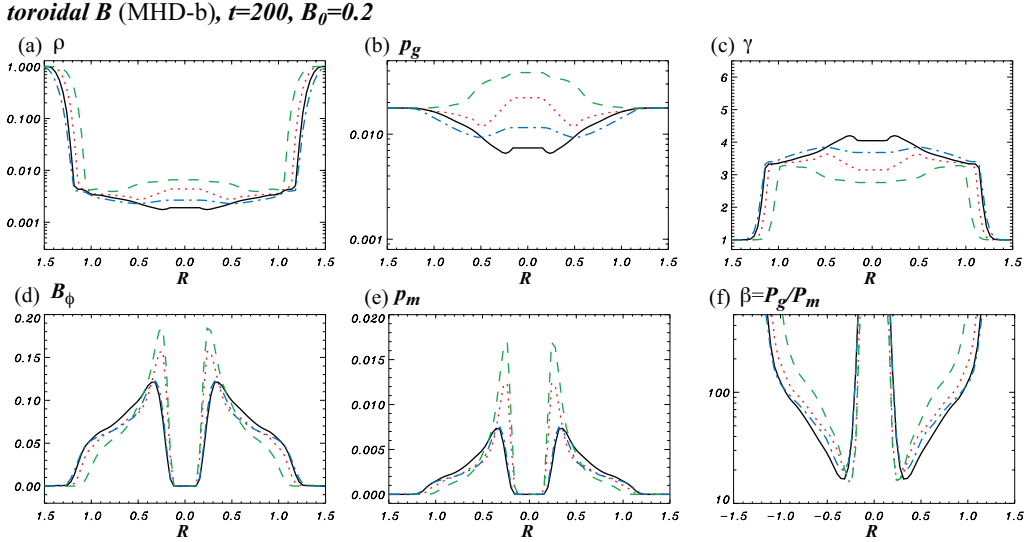}
\end{center}
\caption{The same as in Fig. \ref{1D_onz_axial_comp}, but for the
  toroidal-magnetic field case. \label{1D_toroidal}}
\end{figure*}

The general behavior is similar to both the hydrodynamic and axial
magnetic-field cases. Namely, the over-pressured toroidally magnetized
jet produces an initially weak conical shock that propagates into the
ambient medium and, at the same time, a conical rarefaction wave that
propagates into the jet. The latter expands radially, but the rarefaction
waves produced are also responsible for the conversion of the thermal
energy of the plasma into kinetic energy of the jet following the basic
mechanism of the AR booster. At $z \simeq 4.5$, this process leads to an
increase in the Lorentz factor of flow that is accelerated from $\gamma
\sim 3$ to $\sim 4.2$ (see the solid lines in Fig. \ref{1D_toroidal}c).

Despite these analogies, the toroidal-field case also presents an
important qualitative difference. Although it increases, the maximum
value attained by the jet Lorentz factor is \textit{smaller} than in the
hydrodynamical case; this is in contrast with what was found for the
axial magnetic-field case in the previous Section. The origin of this
different behaviour is to be found in the magnetic tension introduced by
the toroidal magnetic field and that essentially resists both inwards and
outwards motions. This is quite visible in the behaviour along the $z$
direction of the magnetic pressure [see panel \textit{(c)} of
  Fig. \ref{2D_toroidal}], which shows a ``pinching'' effect at $z \sim
4, 13$ and $z \sim 25$, which also correspond to the locations where the
rarefaction waves converge at the jet axis [see panel \textit{(a)} of
  Fig. \ref{2D_toroidal}].

This structure in the magnetic pressure also appears in the radial
profile of the plasma beta in Fig. \ref{1D_toroidal}, which reports 1D
profiles of several quantities perpendicular to the jet at axial
positions $z=4$, $7$, $14$, and $20$. In particular, the panel
\textit{(f)} shows that plasma beta has two minima of the order of
$\beta_p \sim 20$ on either side of the jet axis where the toroidal
magnetic field is at a maximum (\ie at $R \sim \pm 0.3$); this behaviour
should be contrasted with the corresponding one shown in \textit{(f)} of
Fig. \ref{1D_axial} for the axial magnetic-field case, where instead
  the smallest value of the beta parameter is $\beta_p \sim 5$.

\begin{figure}
\begin{center}
\includegraphics[height=10.0cm]{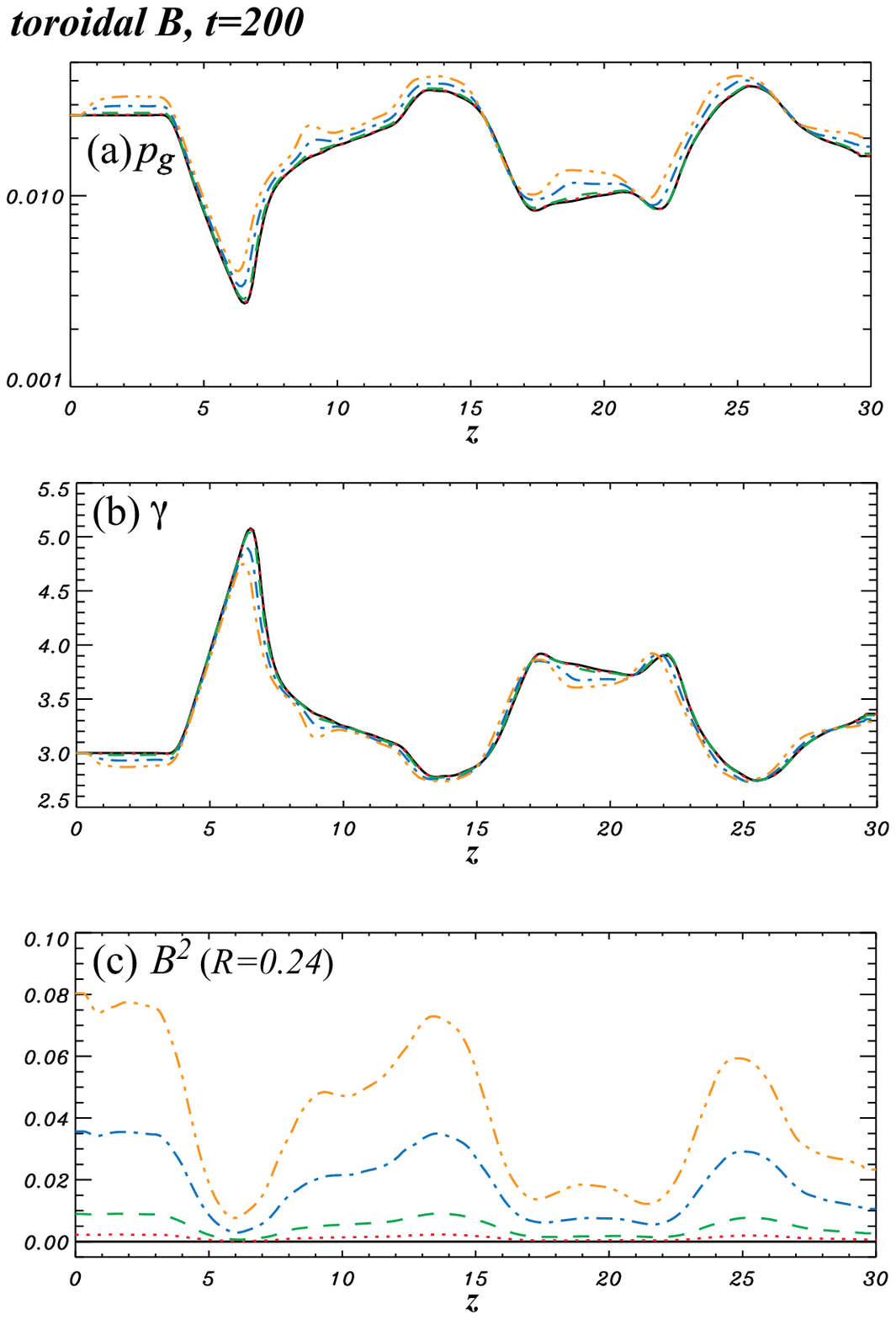}
\end{center}
\caption{The same as in Fig. \ref{1D_onz_axial_comp}, but for the
  toroidal magnetic field case with $B_0=0$ (black solid), $0.05$ (red
  dotted), $0.1$ (green dashed), $0.2$ (blue dash-dotted), and $0.3$
  (orange dash-double-dotted).
\label{1D_onz_toroidal_comp}}
\end{figure}

The presence of a toroidal magnetic field also alters considerably the
flow after the shock encounters the contact discontinuity at $z \simeq
13$, breaking the appearance of a periodicity in the recollimation-shock
structure. Interestingly, both the gas pressure and the Lorentz factor
show an ``O''-shaped structure around $z \simeq 16 - 23$ [see panels
  \textit{(b)} and \textit{(d)} of Fig. \ref{2D_toroidal}], with both of
these quantities reaching a local minimum around $z \simeq 20$ and $R
\simeq 0$ [see also panels \textit{(b)} and \textit{(f)} of
  Fig. \ref{1D_toroidal}].

In summary, the presence of a toroidal magnetic field leads to weaker
recollimation shocks and rarefaction waves, with a more complicated
downstream structures than those found for the hydrodynamic and axial
magnetic field cases. If observed, this different phenomenology should
provide useful information to deduce the properties of the magnetic-field
topology in the jet and, in particular, to establish whether this is
purely toroidal.

Finally, also in this case, we report in Fig. \ref{1D_onz_toroidal_comp}
the 1D profiles of the gas pressure and of the Lorentz factor along the
jet axis, \ie at $R = 0$, as well as the magnetic-field strength along a
direction slightly off the axis, \ie at $R=0.24$ (we recall that the
magnetic field is zero along the axis). The figure aims at establishing
the variations introduced by the initial magnetic-field strength $B_0$
and so different lines refer to $B_0=0.0$ (\ie the purely hydrodynamical
case) as well as to $B_0=0.05$, $0.1$, $0.2$, and $0.3$. Perhaps not
remarkably, only minimal changes are visible in the radial profiles of
the rest-mass density, of the gas pressure and of the jet Lorentz factor
as the magnetic field strength is increased. This is mostly due to the
fact that the magnetic tension introduced by the toroidal magnetic field
acts mostly in the $R$ direction, producing only high-order variations in
the direction of propagation of the jet. Such a behaviour is markedly
different from the one encountered in the case of an axial magnetic
field, and indeed Fig. \ref{1D_onz_toroidal_comp} should be contrasted
with the corresponding Fig. \ref{1D_onz_axial_comp}.

\subsection{Helical magnetic field}
\label{sec:mhd-c}

\begin{figure*}
\begin{center}
\includegraphics[height=7.0cm]{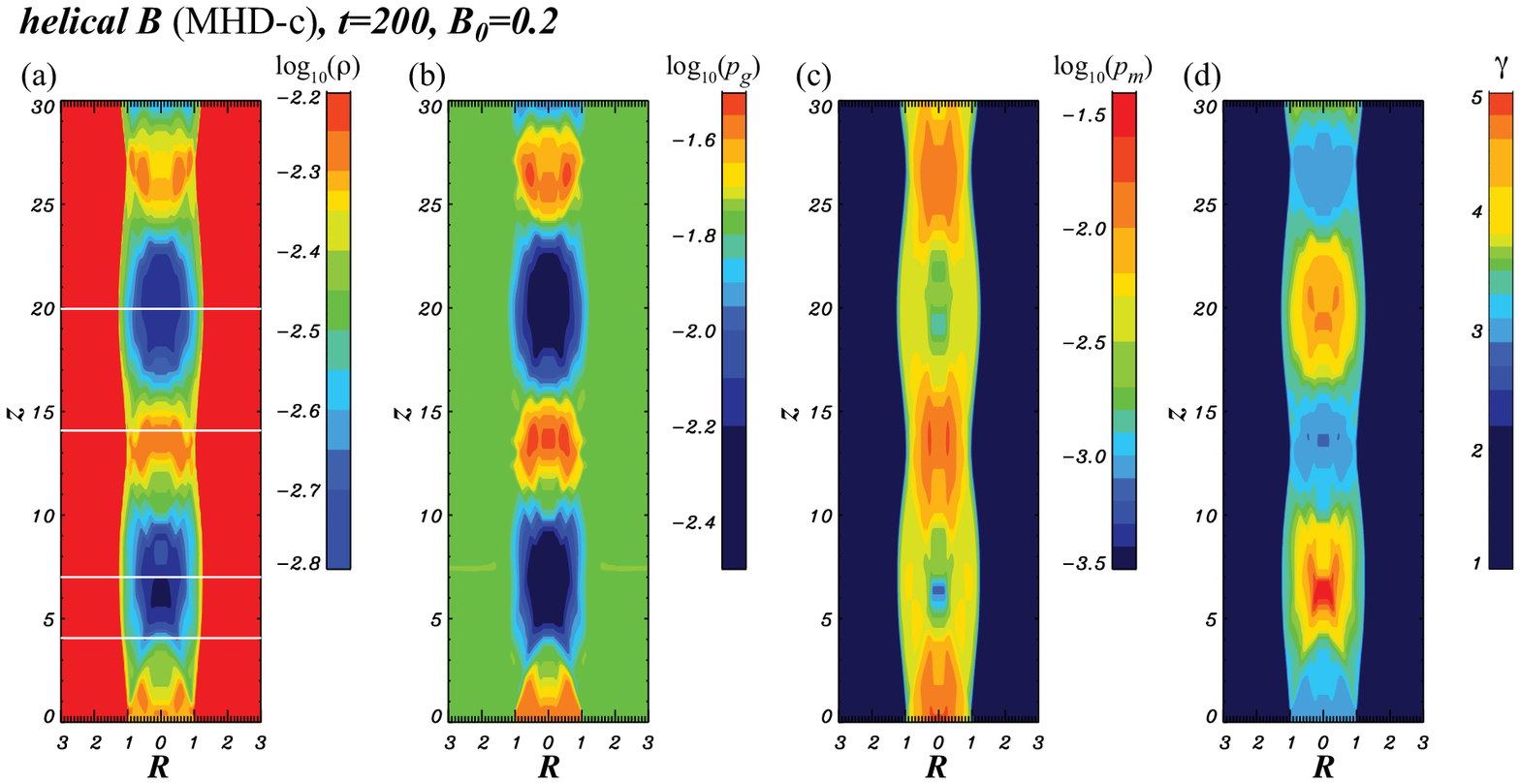}
\end{center}
\caption{Case \texttt{MHD-c}. The same as in Fig. \ref{2D_axial}, but for
  the helical magnetic-field case with $B_0 =0.2$. \label{2D_helical}}
\end{figure*}

\begin{figure*}
\begin{center}
\includegraphics[width=0.75\textwidth]{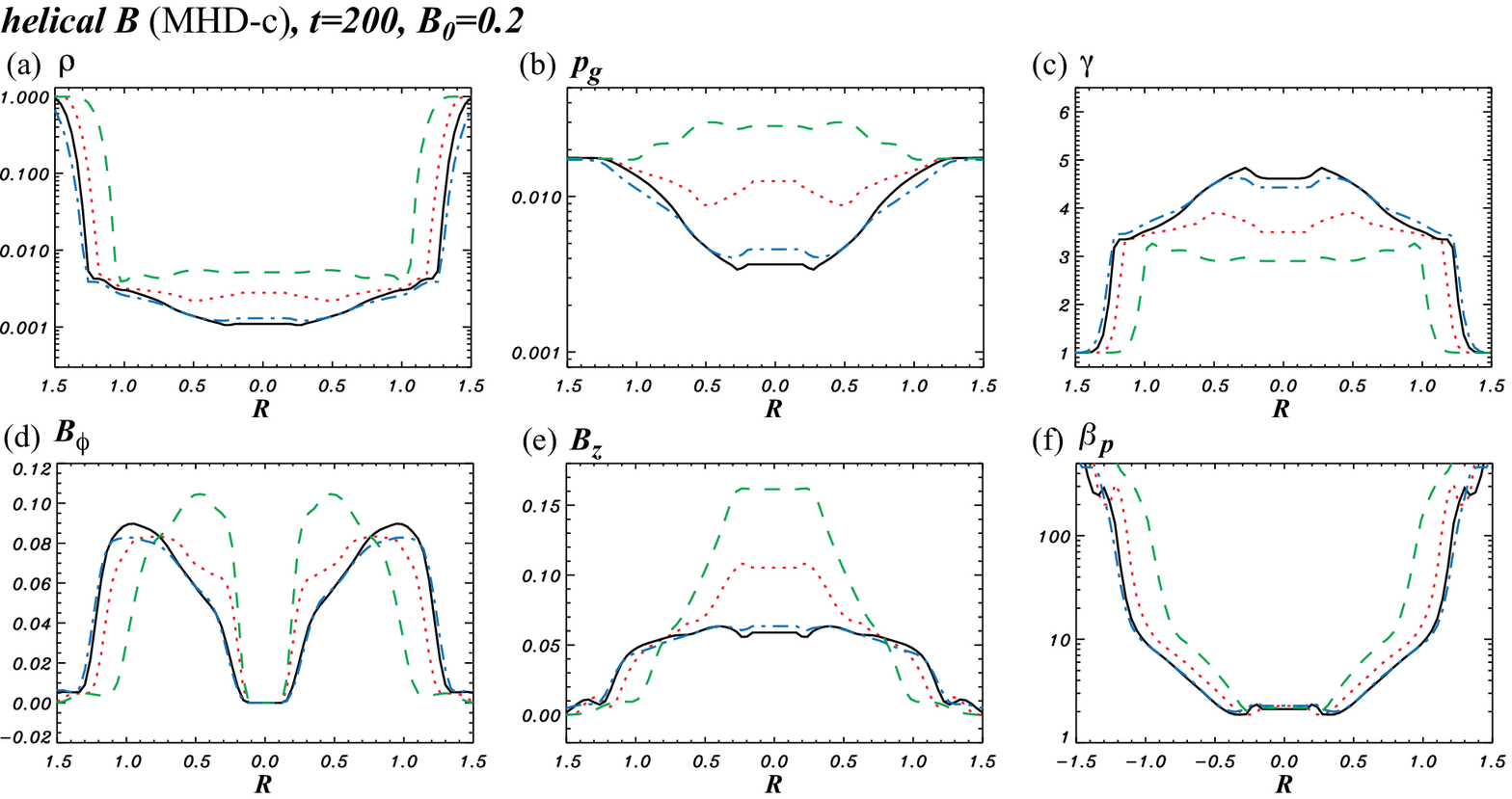}
\end{center}
\caption{1D profiles perpendicular to the jet axis of: ({\it a})
  the rest-mass density, ({\it b}) the gas pressure, ({\it c}) the Lorentz factor,
  ({\it d}) the toroidal magnetic field, ({\it e}) the axial magnetic field, and
  ({\it f}) the plasma beta as measure at $z=4$ (black solid), $7$ (red
  dotted), $14$ (green dashed), and $20$ (blue dash-dotted). All panels
  refer to the helical magnetic-field case with $B_0=0.2$ at $t_s =
  200$. \label{1D_helical}}
\end{figure*}

Finally, we consider the dynamics of the recollimation shocks when a
helical magnetic field is present initially (case \texttt{MHD-c}), with
constant initial pitch $P_0=1/2$. As for the other cases, we show in
Fig. \ref{2D_helical} 2D plots at $t_s=200$ of the rest-mass density, of
the gas- and magnetic pressures, and of the Lorentz factor when
$B_0=0.2$.

Similar to the other magnetic-field topologies, also in this case the
over-pressured jet produces an initially weak conical shock that
propagates into the ambient medium and a conical rarefaction wave that
propagates into the jet. Again, conservation across the rarefaction wave
of $\gamma h$ implies that a conversion of thermal energy to jet kinetic
energy takes place across the rarefaction wave, with a consequent
acceleration of the flow. Although the pitch considered here is less than
one, so that the toroidal magnetic field is larger than the axial one at
least initially, the effective behaviour of the plasma is closer to the
one seen in the case of an axial magnetic field than in the case of a
toroidal magnetic field. In particular, the increase in the Lorentz
factor is \textit{larger} than in the hydrodynamical case, although only
slightly, \ie with downstream Lorentz factors going up to $\gamma \simeq
5.0$. On the other hand, as we have encountered in the previous section
for a purely toroidal magnetic field, the small pitch also implies that
the plasma beta reach its minimum values there [\cf \textit{(f)} in
  Fig. \ref{1D_toroidal}].

Figure \ref{1D_helical} reports the 1D profiles of several quantities
perpendicular to the jet axis at the positions $z=4$, $7$, $14$, and
$20$. Note the close analogies with the corresponding behaviours shown in
Fig. \ref{1D_toroidal} in the case of a purely toroidal magnetic field.
Also here, the convergence of the rarefaction waves towards the jet axis
leads to strong gas and magnetic-pressure gradients, which ultimately
slow the jet expansion. This ceases at $z \simeq 10 - 14$, when weak
shocks return the jet to an over-pressured structure and with a reduced
Lorentz factor and rather different conditions than the initial ones.

In summary, the presence of a helical magnetic field in the jet leads to
a rather complex behaviour in both the recollimation shock and
rarefaction structure. While the flow can be magnetically dominated with
plasma beta of order unity, the Lorentz factor is increased with respect
to the purely hydrodynamical case, mimicking the behaviour seen for
purely axial fields. This global behaviour is clearly influenced in part
by our choice for the initial pitch and we expect that if a lower
magnetic pitch parameter is chosen, corresponding to a more tightly
wrapped helical field, the toroidal field will dominate over the axial
field effects, with a sub-hydrodynamical Lorentz factor boost (see also
the discussion in Section \ref{se:ja}).

\begin{figure}
\begin{center}
\includegraphics[height=10.0cm]{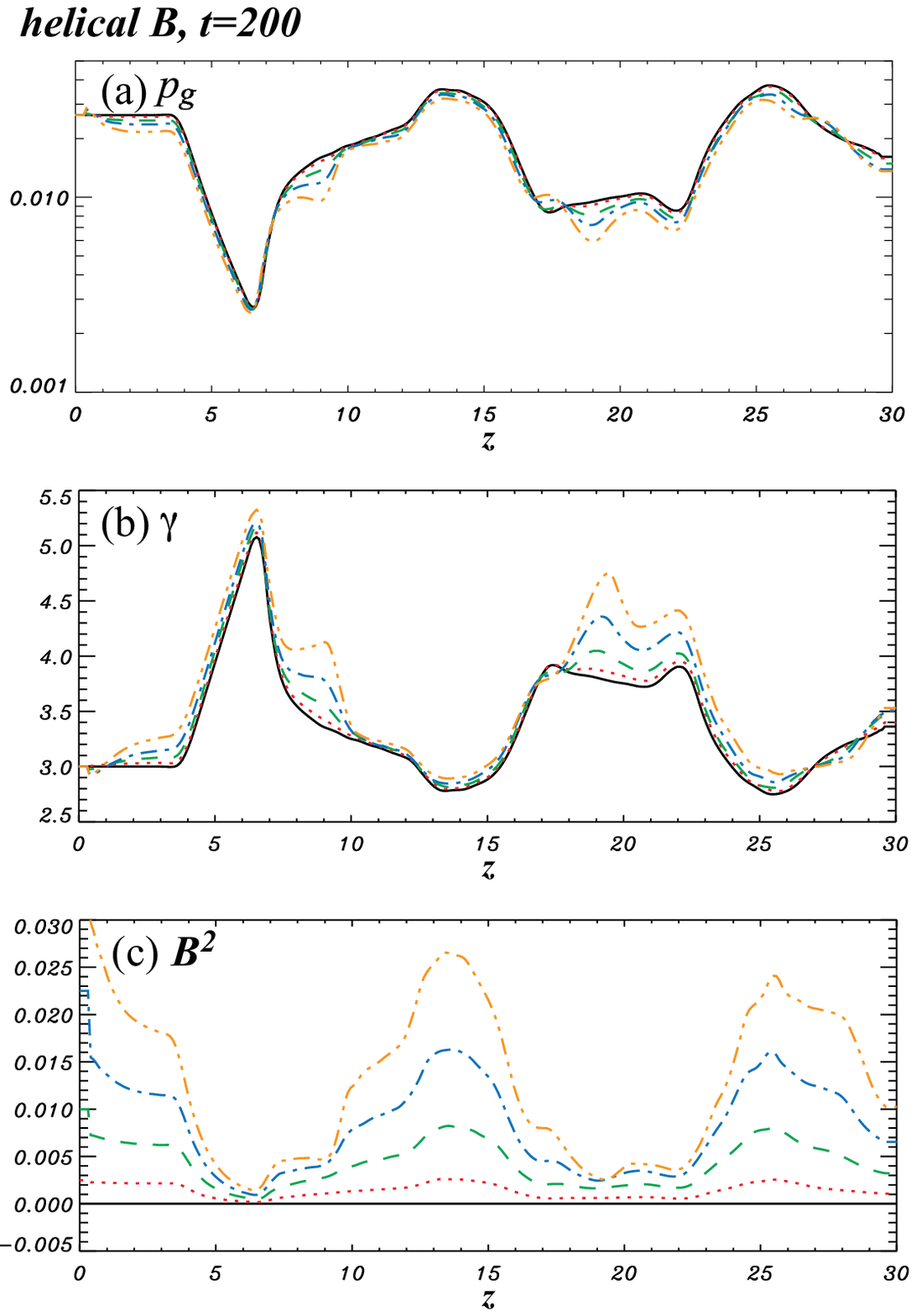}
\end{center}
\caption{The same as in Fig. \ref{1D_onz_axial_comp}, but for the helical
  magnetic-field case with $B_0=0$ (black solid), $0.05$ (red dotted),
  $0.1$ (green dashed), $0.15$ (blue dash-dotted), and $0.2$ (orange
  dash-double-dotted).
\label{1D_onz_helical_comp}}
\end{figure}

Finally, the role played by the initial strength of the magnetic field
$B_0$ on the subsequent dynamics is shown in
Fig. \ref{1D_onz_helical_comp}, which reports the 1D profiles along the
jet's axis of the gas pressure, of the Lorentz factor and of the magnetic
field strength. Again, the different lines refer respectively to
$B_0=0.0$, and to $B_0=0.05$, $0.1$, $0.15$, and $0.2$ (dotted, dashed,
dash-dotted and dash-double-dotted lines, respectively). Also in this
case, it is possible to observe how the phenomenology of the helical
field is a combination of the one seen for purely axial and purely
toroidal magnetic fields. In particular, as in the toroidal-field case,
the gas pressure shows only modest changes along the $z$ direction when
the magnetic field is increased. On the other hand, as in the axial-field
case, the Lorentz factor shows peaks that increase and move
downstream with increasing magnetic-field strengths.

\section{Jet Acceleration}
\label{se:ja}

%
\begin{figure*}
\begin{center}
\includegraphics[width=0.95\textwidth]{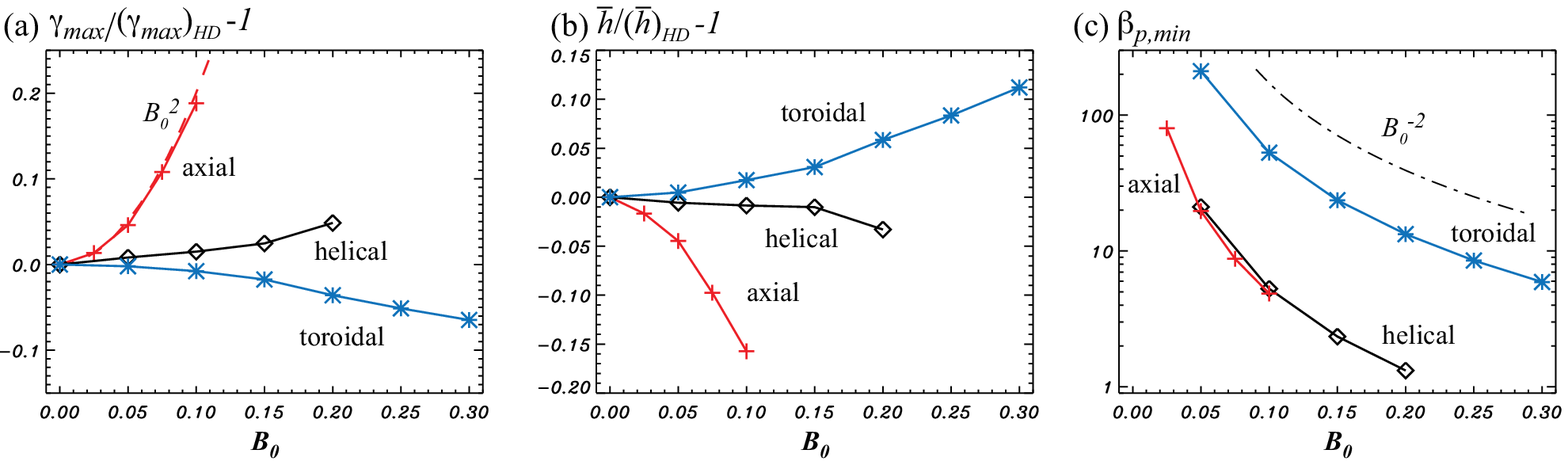}
\end{center}
\caption{Dependence of: ({\it a}) the relative difference of the maximum
  Lorentz factor $\gamma_{\rm max} / (\gamma_{\rm max})_{_{\rm HD}}-1$,
  ({\it b}) the relative difference of the specific enthalpy
  $\bar{h}/(\bar{h})_{_{\rm HD}} - 1$, and ({\it c}) the minimum plasma
  beta. All quantities are shown as a function of the initial
  magnetic-field strength, $B_0$ in the case of axial (red crosses),
  toroidal (blue stars), and helical magnetic fields (black
  diamonds) with $P_0=0.5$. The red dashed line in panel ({\it a}) indicates a
  quadratic fit [\cf Eq. \eqref{eq:fit_quad}], while the black dot-dashed
  line in panel ({\it c}) refers to a $B_0^{-2}$ scaling.
\label{dep1}}
\end{figure*}

In the previous Sections we have illustrated in detail how a robust
feature of the recollimation shocks considered here is the acceleration
of the flow downstream of the inlet as a result of the AR booster. We
have also remarked that the occurrence of this acceleration is the result
of the conversion of the plasma thermal energy into kinetic energy of the
jet, or, more precisely, of the conservation of the quantity $\gamma h$
across a rarefaction wave \citep{Alo06}. In addition, we have discussed
how different magnetic field topologies lead to systematically different
values of the maximum increase in the Lorentz factor measured downstream
of the inlet, \ie $\Delta \gamma_{\rm max} \equiv {\rm max}({\gamma}) -
\gamma_0$. What we have not yet discussed, however is how this
accelerating mechanism depends on our choice of the initial
magnetic-field strength, $B_0$.

\begin{figure*}
\begin{center}
\includegraphics[width=0.95\textwidth]{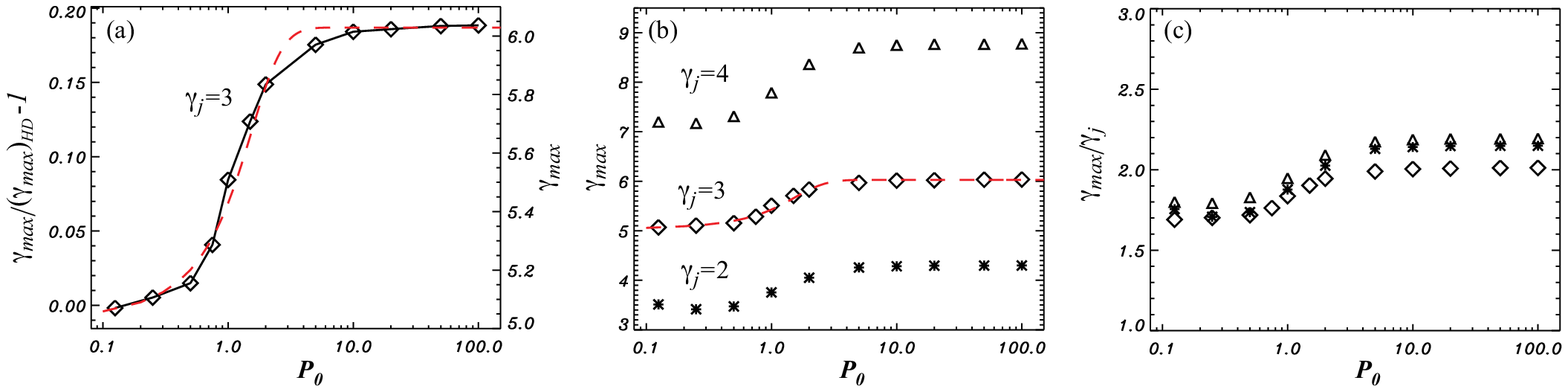}
\end{center}
\caption{\textit{Left panel:} Dependence of the maximum Lorentz
    factor $\gamma_{\rm max}$ (right vertical axis) and relative
    difference of the maximum Lorentz factor relative to the purely
    hydrodynamical case $\gamma_{\rm max}/(\gamma_{\rm max})_{_{\rm HD}}
    -1$ (left vertical axis) as a function of the initial magnetic pitch,
    $P_0$; the initial jet Lorenz factor is set to $\gamma_j=3$; the red
    dashed line indicates the fitting with a hyperbolic tangent function
    [\cf Eq. \eqref{eq:pitch_dep}]. \textit{Central panel:} Dependence of
    the maximum Lorentz factor as a function of the initial magnetic
    pitch, for three different values of the initial jet Lorenz factor
    $\gamma_j=2$ (stars), $3$ (diamonds), and $4$ (triangles); the red
    dashed line indicates the fitting as in the left panel. \textit{Right
      panel:}. The same as in the central panel but when the maximum
    Lorentz factors are normalized to the initial one; note that in this
    case all curves essentially overlap. All cases refer to an initial
    magnetic field $B_0 = 0.1$. \label{dep2}}
\end{figure*}
\vskip 4.0cm

This point is addressed in the left panel of Fig. \ref{dep1}, which
reports the relative increase of the Lorentz factor with respect to the
purely hydrodynamical evolution, namely, $\Delta \gamma_{\rm max}/
(\gamma_{\rm max})_{_{\rm HD}} \equiv \gamma_{\rm max}/(\gamma_{\rm
  max})_{_{\rm HD}} - 1$, as a function of the initial magnetic
field. Obviously, this quantity can either be positive or negative and
provides a direct measure of the fractional boost.  Shown with different
symbols are the different magnetic-field topologies, with crosses
referring to the axial magnetic field (case \texttt{MHD-a}), diamonds to
the helical magnetic-field (case \texttt{MHD-c}), and star crosses to the
toroidal magnetic-field (case \texttt{MHD-b}).

When presented in this manner, it is then straightforward to realize that
axial and helical initial magnetic fields lead to Lorentz boosts that are
larger than in the hydrodynamical case, while the opposite is true for
purely toroidal magnetic fields, for which an acceleration is still
present but this is smaller than in the hydrodynamical case. We have
already discussed in the previous Sections that the origin of this
different behaviour has to be found in the fact that an axial magnetic
field adds an effective gas pressure and results in larger rest-mass
density and pressure gradients across the rarefaction waves induced
downstream of the inlet. In turn, the AR booster translates these
stronger waves into larger accelerations of the flow. It is also
instructive that, in the case of a purely axial magnetic field, the
behaviour of the relative boost has a simple quadratic dependence on the
initial magnetic field. This is simply because the relative boost scales
as
\begin{equation}
\label{eq:fit_quad}
\frac{\Delta \gamma_{\rm max}}{(\gamma_{\rm max})_{_{\rm HD}}} 
\propto \frac{p}{p_g} - 1 = \frac{p_m}{p_g} \propto B_z^2\,.
\end{equation}
This is confirmed by the very good match between the numerical data and
a quadratic fit, which is indicated with a red dashed line.

Also shown in Fig. \ref{dep1}, but in the middle panel, are the
fractional differences in the specific enthalpy at the position of the
maximum Lorentz factor, \ie $\bar{h}/(\bar{h})_{_{\rm HD}} - 1$. Here
too, different symbols refer to the various magnetic-field topologies
(\cf left panel) and different initial strengths. Clearly, this panel
offers a complementary view to the one in the left panel and summarizes
much of what already discussed in the previous Sections. Namely, that
axial and helical magnetic fields leads to smaller values of the specific
enthalpy in the downstream solution, in contrast to the case of toroidal
magnetic fields where instead the values of the specific enthalpy at
Lorentz-factor maximum increases with the initial magnetic
field. Finally, shown in the right panel of Fig. \ref{dep1} is the
behaviour of the minimum plasma beta $\beta_{p, {\rm min}}$ as a function
of the initial magnetic field. Interestingly, all magnetic-field
topologies show the same (and expected) quadratic dependence as
$B^{-2}_0$.

\subsection{Dependence on magnetic pitch}

As discussed in Section \ref{sec:ns}, in the case of a helical magnetic
field, we have an additional degree of freedom represented by the initial
magnetic pitch as defined in Eq. \eqref{eq:mp}. By suitably choosing the
initial pitch, \ie the ratio $a/R_j$, it is possible to scan the range of
possible magnetic-field configurations, which range from an essentially
toroidal magnetic field\footnote{Note that in practice even if the
  magnetic pitch is very small, the magnetic field does not reach the
  toroidal magnetic-field profile discussed in Section \ref{sec:mhd-b}
  because there is always a nonzero poloidal component at the jet center
  $R=0$.}  for $P_0 \ll 1$ to an axial axial magnetic field for $P_0 \gg
1$.

Given this freedom in the parameterization, and given that different
magnetic-field topologies lead to different amplifications to the AR
booster, it is interesting to ask whether a correlation exists between
the initial pitch and the increase in the maximum Lorentz factor
downstream of the inlet. This is shown in Fig. \ref{dep2}, which shows
$\Delta \gamma_{\rm max}/ (\gamma_{\rm max})_{_{\rm HD}}$ as a function
of the initial pitch $P_0$ for a number of simulations carried out with
$B_0=0.1$. Interestingly, we found that the relative increase in the
maximum Lorentz factor has a very clear dependence with the pitch,
smoothly joining the two extreme cases of a toroidal and axial magnetic
fields, respectively. Furthermore, the transition between the two regimes
takes place at $P_0 \gtrsim 1$, that is, when $a \gtrsim R_j$, saturating
to the axial field case when $a \simeq 10\, R_j$. Finally, the dependence
can be accurately approximated with a simple expression of the type
\begin{equation}
\label{eq:pitch_dep}
\gamma_{\rm max} 
\simeq c_1 + c_2 \tanh[c_3(P_0-1.0)]\,,
\end{equation}
where $c_1 \simeq 5.43, c_2 \simeq 0.6$, and $c_3 \simeq 0.8$, 
and is indicated with the red dashed line in Fig. \ref{dep2}.

The relevance of expression \eqref{eq:pitch_dep} is that it provides,
  at least in principle, yet another useful tool to deduce the properties
  of the magnetic-field topology from the observations of the jet
  dynamics. Indeed, if radio-astronomical observations could provide a
  reliable measure of the Lorentz factor in the bulk of the jet and at
  its base, \ie $\gamma_{\rm max}$ and $\gamma_j$, then expression
  \eqref{eq:pitch_dep} would provide a simple way to deduce the degree of
  helicity of the magnetic field in the jet. This is because while the
  relation between $\gamma_{\rm max}$ and the pitch $P_0$ changes with
  the initial Lorentz factor in the jet $\gamma_j$, the fact that
  $\gamma_{\rm max}$ and $\gamma_j$ scale linearly [\cf left panel of
    Fig. \ref{hydro_dep}] implies that the functional dependence of
  $\gamma_{\rm max}/\gamma_j$ with $P_0$ will not change. This is shown
  in the right panel of Fig. \ref{dep2}. Hence, once and if $\gamma_{\rm
    max}/\gamma_j$ is measured, a direct estimate on $P_0$ is possible,
  which is only weakly dependent on the jet overpressure. Conversely, if
  the magnetic field pitch angle could be determined, \eg from
  polarimetric observations, then it could be possible to estimate
  $\gamma_{\rm max}$. Such a measurement would obviously be of great
  importance for the interpretation of the very rapid TeV variability in
  AGN jets.

\section{Conclusion}

We have performed 2D SRMHD simulations of the propagation of an
over-pressured relativistic jet leading to the generation of a series of
recollimation shocks and rarefaction waves. The overall dynamics of this
process is rather well known.  Downstream of the inlet, the jet produces
a weak conical shock that propagates into the ambient medium and a
quasi-stationary conical rarefaction wave that propagates inwards. The
significant drops in rest-mass density and pressure produced by the
rarefaction waves are also responsible for the conversion of the thermal
energy into kinetic energy of the jet \citep[see, \eg][]{Alo06, Mat12}.
This is a purely relativistic effect that develops in the presence of
strong tangential discontinuities in the flow \citep{Rez02}, and that
leads to a consistent and robust boost of the fluid. In addition, the
nonlinear interaction of the shocks and rarefaction waves lead to a
stationary and multiple recollimation-shock structure along the jet,
which has been reproduced by a number of the hydrodynamic simulations
under different physical conditions \citep[see,
  \eg][]{Gom97,Kom97,Agu01,Alo03,Roc08,Roc09,Mim09,Mat12}.

We have here extended previous hydrodynamic work to determine the effect
that magnetic fields with axial, toroidal, and helical topologies have on
the jet dynamics. In particular, we have found that an axial magnetic
field behaves as an additional gas pressure, leading to a sharper
recollimation-shock structure and thus larger flow accelerations. On the
other hand, a toroidal magnetic field tends to reduce the inward motion
of the rarefaction waves, thus leading to a weaker recollimation-shock
structure and accelerations that are even smaller than in the case of a
purely hydrodynamical evolution. Finally, a helical magnetic field tends
to yield a behaviour which is a combination of those observed for purely
axial/toroidal magnetic fields, with larger Lorentz factors, but also
with a rather complex recollimation-shock substructure.

As predicted by the basic properties of the AR booster, we have found
that the boost in the jet flow, as measured in terms of the maximum
Lorentz factor, is anti-correlated with the values of the specific
enthalpy. Furthermore, the boost grows quadratically with the initial
magnetic field in the case of a purely axial flow, at smaller rates in
the case of a helical magnetic field, and it decreases systematically
when a purely toroidal magnetic field is present. Finally, we have shown
that the maximum Lorentz factor exhibits a smooth behaviour in terms of
the initial pitch in the case of an axial magnetic field.

Stationary components are commonly seen in parsec-scale VLBI observations
of AGN jets \citep[see, \eg][]{Jor05, Lis13, Coh14}, being normally
associated with recollimation shocks. Studying in detail the structure in
these stationary components for a direct comparison with our simulations
requires however resolving the jet structure across the jet width. This
can be achieved through ``regular'' cm-VLBI observations of nearby
sources, like M87, but for the majority of the AGN jets it would require
a significant increase in angular resolution. This can be obtained
through VLBI observations at even shorter wavelengths, in which the 3\,mm
GMVA (Global Millimeter VLBI Array) and 1.3\,mm VLBI observations with
the Event Horizon Telescope and Black-Hole Cam projects can achieve
angular resolutions between 50 and 20 microarcseconds \citep[see,
  \eg][]{Fis14, Kri13}, and through space VLBI observations. The space
VLBI mission ``RadioAstron'' has recently successfully achieved
ground-space fringe detections for observations at 1.3\,cm on baselines
longer than eight Earth diameters, allowing imaging the innermost regions
in AGN jets with an unprecedented angular resolution of $\sim$ 20
microarcseconds \citep{Gom15}.

Whilst the range of simulations carried out here have tried to explore a
rather large portion of the space of parameters, a number of improvements
on the approach followed here can be made. First, even though we have
assumed an over-pressured jet to produce the recollimation-shock
structure, a pressure mismatch between jet and ambient medium arises
naturally if the pressure in the ambient medium decreases with distance
from the central object. In this case, the recollimation-shock structure
depends on the ambient-medium pressure scale \citep[see,
  \eg][]{Gom97,Kom97,Agu01,Alo03,Mim09,Koh12a,Mat12,Por14}. Although we
do not expect that significant qualitative differences will emerge from a
more realistic modelling of the ambient medium, we will extend the
current investigation to include larger overpressure ratios, as well as
declining pressure profiles outside the jet.

Second, in this work we assume the relativistic jet is kinematically
dominated initially for all values of the initial magnetic
field. However, magnetically dominated relativistic jets (Poynting-flux
dominated jets) might be plausible from the results of previous
analytical \citep[\eg][]{Hey03, Bes09, Lyu09} and numerical studies
\citep[\eg][]{Kom06, Tch09, Por11, McK12}. We will investigate this type
of relativistic jets in future work.

Third, it is clear that the magnetic modifications of the
recollimation-shock structure found here may affect in a significant way
the signatures of stationary components seen in several AGN jets, 
especially in polarized flux. An even
clearer signature difference may be obtained when a shock like
disturbance propagates through the stationary recollimation-shock
structure. In view of this, we will investigate the propagation of such a
shock-like disturbance through a magnetized recollimation shock and
calculate the corresponding emission following the approach presented in,
\eg \citet{Gom97}.

Finally, although our simulations show that the recollimation-shock
strength depends on the jet magnetization and magnetic field structure,
it is clear that its observed appearance as imaged by VLBI would also
depend on the jet-viewing angle, and is obviously influenced by
Doppler-boosting effects. Hence, depending on the jet bulk flow Lorentz
factor and viewing angle, the Doppler boosting may cancel, or even
reverse, the increased emissivity obtained in the recollimation shock and
due to the enhanced particle and magnetic-field energy rest-mass density
\citep[see, \eg][]{Gom97,Alo03,Roc08, Roc09, Mim09}. These effects will
also be the focus of future investigations.

\acknowledgments
Support comes the ERC Synergy Grant ``BlackHoleCam -- Imaging the Event
Horizon of Black Holes'' (Grant 610058) and from the Ministry of Science
and Technology of Taiwan under the grant NSC 100-2112-M-007-022-MY3 and
MOST 103-2112-M-007-023-MY3. J. L. G. acknowledges support from the
Spanish Ministry of Economy and Competitiveness grant AYA2013-40825-P.
K. N. and P. H. acknowledge support by NSF awards AST-0908010 and
AST-0908040, and by NASA awards NNX09AD16G, NNX12AH06G, NNX13A P-21G, and
NNX13AP14G. A. M. acknowledges support from the Scientific Research (FWO)
and the Belgian Federal Science Policy Office (Belspo). The simulations
were performed on Pleiades at NASA, on SR16000 at Kyoto University, on
Nautilus at the University of Tennessee, and on LOEWE at the Goethe
University Frankfurt.

\newpage
\appendix

\section{Convergence Tests}
\subsection{Large-scale hydrodynamical jet}
\label{se:hdlong}

\begin{figure*}
\begin{center}
\includegraphics[width=0.7\textwidth]{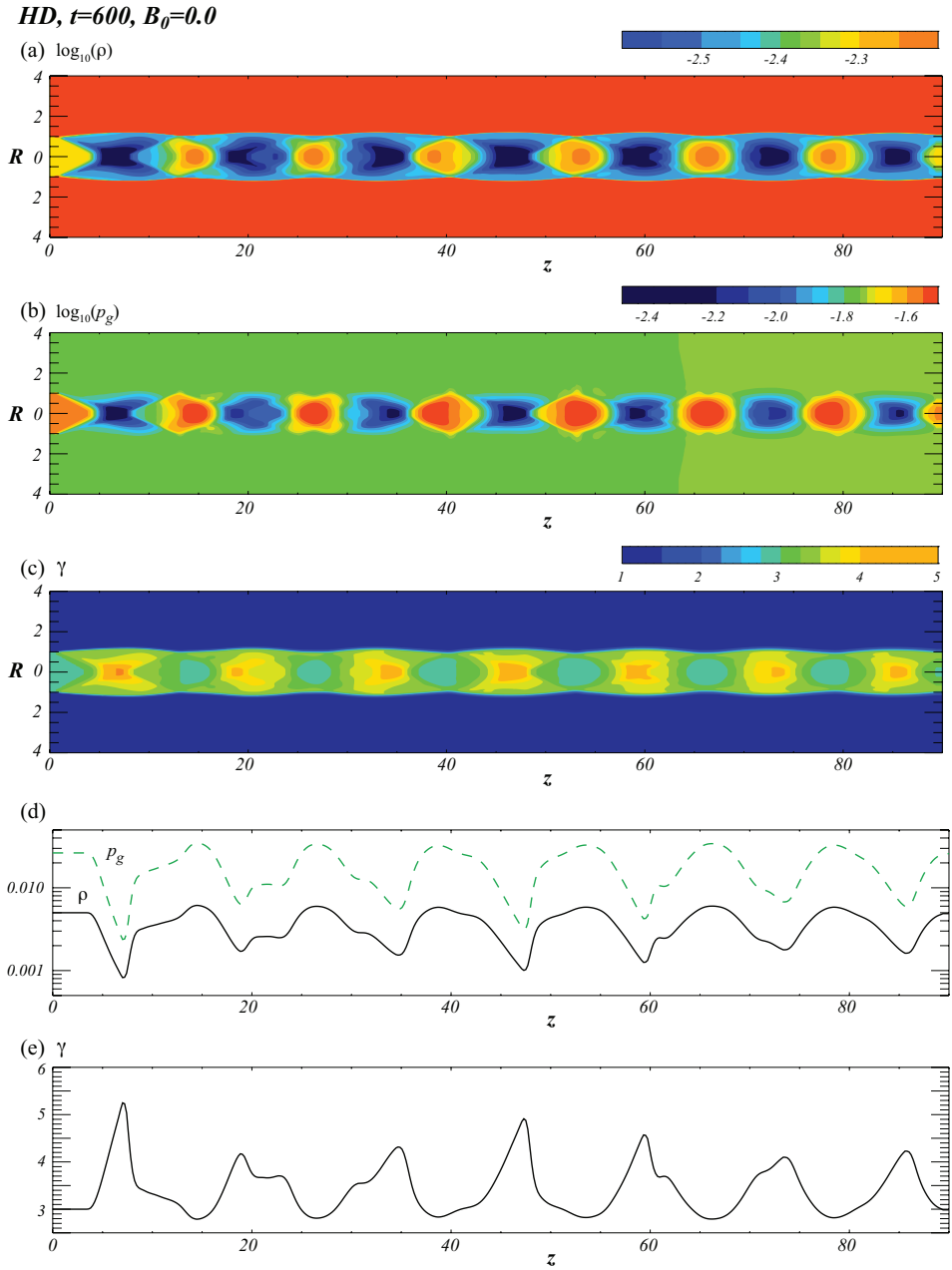}
\end{center}
\caption{{\it Upper panels:} 2D plots of: ({\it a}) the rest-mass
  density, ({\it b}) the gas pressure, and ({\it c}) the Lorentz factor.  {\it
    Lower panels:} 1D profiles along the jet axis ($R=0$) of: ({\it
    d}) the rest-mass density (solid) and the gas pressure (green dashed), and
  ({\it e}) the Lorentz factor. All panels refer to $t_s = 600$ for a purely
  hydrodynamical jet and an extent in the $z$ direction that is three
  times larger than the one considered in the main text.
\label{2D_hydro_long}}
\end{figure*}

 Due to the uniform ambient medium, the modifications of the
  recollimation-shock and rarefaction-wave structures is expected to be
  rather small and a quasi-periodic structure should develop on larger
  axial scales \citep[see also][]{Mat12}. In this Appendix we test this
  assumption, and hence the development of a self-similarity in the
  recollimation-shock structure, by considering a purely hydrodynamical
  simulation having the same resolution discussed in Section
  \ref{sec:hd}, but with an extent in the $z$ direction from $0$ to $90$,
  that is, three times larger than what presented in the main text. The
  results of this simulation are reported in Fig. \ref{2D_hydro_long},
  with the top three panels referring to a two-dimensional view, and the
  bottom two panels to a cut along the $z$ axis at $t_s=600$. It is clear
  that the shock/rarefaction structure is periodic and with periodicity
  $\simeq 12 R_j$. This separation distance depends on the initial jet
  velocity, over-pressure ratio between jet and ambient pressure, and
  opening angle of the jet.
%

\subsection{Dependence on initial jet Lorentz factor and over-pressure ratio 
of the maximum jet Lorentz factor}
\label{se:hddep}

\begin{figure*}
\begin{center}
\includegraphics[width=10.0cm]{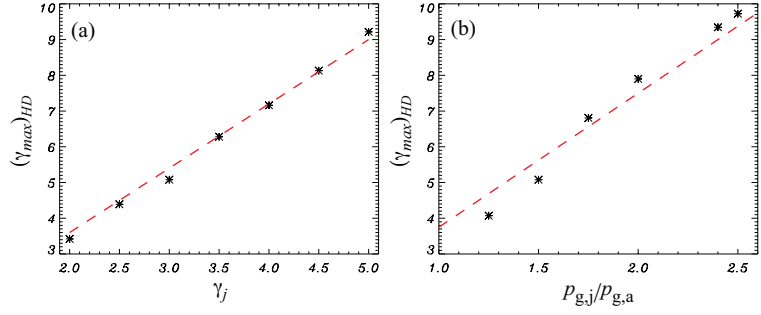}
\end{center}
\caption{Dependence of the maximum Lorentz factor in a purely
  hydrodynamical jet $(\gamma_{jet})_{_{\rm HD}}$ as a function of: ({\it
    a}) the initial jet Lorentz factor $\gamma_j$ with fixed
  over-pressure ratio $p_{g,j}/p_{g,a}=1.5$ and ({\it b}) the
  over-pressure ratio $p_{g,j}/p_{g,a}$ with fixed initial jet Lorentz
  factor $\gamma_j=3$. The red dashed lines indicate the linear fitting.
\label{hydro_dep}}
\end{figure*}
 As discussed in Section \ref{se:ja}, the relative difference of the
  maximum jet Lorentz factor with respect to the purely hydrodynamic case
  depends on the initial magnetic field strength and magnetic pitch.  It
  is important to know the maximum jet Lorentz factor of a pure
  hydrodynamic case.  In this Appendix, we have investigated the
  dependence of the maximum jet Lorentz factor of a pure hydrodynamic jet
  on the initial jet Lorentz factor and over-pressure ratio.  Figure
  \ref{hydro_dep} shows $(\gamma_{max})_{_{\rm HD}}$ as a function of the
  initial jet Lorentz factor ($\gamma_j$) and of the over-pressure ratio
  ($p_{g,j}/p_{g,a}$). Clearly, $(\gamma_{jet})_{_{\rm HD}}$ has a simple
  linear dependence on both the initial jet Lorentz factor and the
  over-pressure ratio with a coefficient of $\simeq 1.8$ and $\simeq
  3.75$ respectively, which we indicate with the red dashed lines in
  Fig. \ref{hydro_dep}.

\end{document}